\documentclass[preprint2]{aastex}

\usepackage{bm}

\shorttitle{Photoevaporation of Circumstellar Disks}
\shortauthors{Tanaka et al.}

\begin{document}

\title{Photoevaporation of Circumstellar Disks Revisited:\\
The Dust-Free Case}

\author{Kei E. I. Tanaka\altaffilmark{1,2,3} , Taishi Nakamoto\altaffilmark{3}, and Kazuyuki Omukai\altaffilmark{1,2}}
\affil{1: Astronomical Institute, Tohoku University, Sendai 980-8578, Japan}
\affil{2: Department of Physics, Kyoto University, Kyoto 606-8502, Japan}
\affil{3: Department of Earth and Planetary Sciences, Tokyo Institute of Technology, Tokyo 152-8551, Japan}
\email{ktanaka@astr.tohoku.ac.jp}

\begin{abstract}
Photoevaporation by stellar ionizing radiation 
is believed to play 
an important role in the dispersal of disks around young stars.
The mass loss model for dust-free disks developed by Hollenbach et al. 
is currently regarded as a conventional one and has been used in a
wide variety of studies.
However, the rate in this model was derived by the crude so-called 1+1D
approximation of ionizing radiation transfer, which assumes 
that diffuse radiation propagates in a direction vertical to the disk.
In this study, we revisit the photoevaporation 
of dust-free disks by solving the 2D axisymmetric radiative 
transfer for steady-state disks. 
Unlike that solved by the conventional model, 
we determine that direct stellar radiation is more important 
than the diffuse field at the disk surface. 
The radial density distribution at the ionization 
boundary is represented by the single power-law with an index $-3/2$
in contrast to the conventional double power-law.
For this distribution, the photoevaporation rate from the entire disk can be written
as a function of the ionizing photon emissivity $\Phi_{\rm EUV}$ 
from the central star and the disk outer radius $r_{\rm d}$ as follows:
$\dot{M}_{\rm PE} = 5.4 \times 10^{-5}
	(\Phi_{\rm EUV}/10^{49}{\rm sec}^{-1})^{1/2}
	(r_{\rm d}/1000{\rm AU})^{1/2}
	M_\odot {\rm yr}^{-1}$.
This new rate depends on the outer disk radius rather than on 
the gravitational radius as in the conventional model,
caused by the enhanced contribution to the mass loss from the outer disk annuli. 
In addition, we discuss its applications to present-day as well as primordial star formation.
\end{abstract}


\section{Introduction}
Stars are formed by the gravitational collapse of 
pre-stellar cores with non-zero angular momentum.
As a natural results, disks are formed around newborn stars,
and most of the materials are 
accreted through thses. 
The final stellar mass at the time of its formation is set when the disk dissipates.
Moreover, disk dissipation determines the formation environments of planets, which are formed inside disks at
the final stage of low-mass star formation.
Photoevaporation is currently considered to be a
promising dissipation mechanism by which the disk gas escapes from the gravitational binding of a star as a result of heating by ultraviolet (UV) radiation from the star 
or external sources.
%{\bf
While the irradiation by nearby stars can be important
in the case of low-mass star formation in a dense cluster,
\citep[e.g.,][]{joh98, ada04, fat08,hol11,tho13},
the photoevaporation by radiation from the central star 
should dominate in isolated or massive star formation.  
In this study, we focus on the latter process.
%}

The photoevaporation of protoplanetary disks around low-mass stars has been studied since 1990s.
Early studies concentrated on photoevaporation by extreme ultraviolet (EUV), i.e., ionizing radiation 
\citep{shu93,cla01} from the central star.
However, photoevaporation by far-ultraviolet (FUV) radiation and/or 
X-rays is now considered as a more dominant mechanism for disks 
around low-mass stars \citep{gor09,owe12} on the basis that its 
timescale to operate, $\sim 3~{\rm Myr}$, is consistent 
with the observational disk-dissipation timescale.
However, this theory remains in dispute.

EUV photoevaporation has attracted significant attention recently in
the context of primordial star formation 
in the early universe.
Although primordial stars were first speculated 
to be $\sim1000~M_\odot$ because of their natal pre-stellar core masses \citep{bro99, omu03},
recent studies that consider stellar feedback onto accretion flows via the disk
demonstrated that the EUV effect on the infalling gas
becomes significant for masses greater than $10~M_\odot$, and that 
photoevaporation terminates mass accretion onto the newborn star at a mass of $20$ -- $100~M_\odot$ 
\citep{mck08, hos11,hos12b,sta12}.
A similar mechanism can be applied to present-day massive star ($>20~M_\odot$) formation, 
although this topic has not been studied in depth thus far.

In studies of 
disk evolution by EUV photoevaporation, 
the formula derived by \citet{hol94} 
(hereafter, HJLS94) has often been used.
By calculating the approximate radiative transfer (RT) 
for assumed steady-state density distributions around circumstellar disks, 
HJLS94 derived the mass-loss rate from dust-free disks as
\begin{eqnarray}
	\dot{M}_{\rm PE, H94} = 1.3 \times 10^{-5}
	is equal to\left (\frac{\Phi_{\rm EUV}}{10^{49}{\rm s}^{-1}} \right)^{1/2} \nonumber \\ 
	\times \left (\frac{M_*}{10M_\odot} \right)^{1/2}
	M_\odot {\rm yr}^{-1},
	\label{eq_pe_rate_h94}
\end{eqnarray}
where $\Phi_{\rm EUV}$ is the EUV photon emission rate, 
and $M_*$ is the mass of the central star.
Although HJLS94 emphasized the importance of 
diffuse radiation, i.e., the re-emitted radiation 
from the ionized atmosphere above the disk, 
in ionizing the disk surface, 
such radiation was not adequately treated in their calculation.
To save computational expence, they adopted the 1+1D approximation for RT, 
where diffuse radiation is assumed to propagate in a direction vertical to the disk.
In reality, of course, diffuse radiation 
also has radial components, because the density and 
ionization degree have radial gradients 
in the atmosphere.
In this study, to accurately treat diffuse radiation, we calculate the axisymmetric 2D RT and re-examine the results of HJLS94.

In Section 2, we describe the basics of 
photoevaporation and our model for calculation. 
In Section 3, we present our results of the 2D 
RT calculation, which is significantly differs from the previous 1+1D result.
In Section 4, we analytically interpret the numerical results and derive a new formula for the photoevaporation rate 
for dust-free disks.
Section 5 is reserved for a discussion on our model's 
implication on star formation and its validity and limitations.
Finally, we summarize our study in Section 6.

%%%%%%%%%%%%%%%%%%
\section{Model}
\begin{figure}[t]
%\epsscale{.60}
\plotone{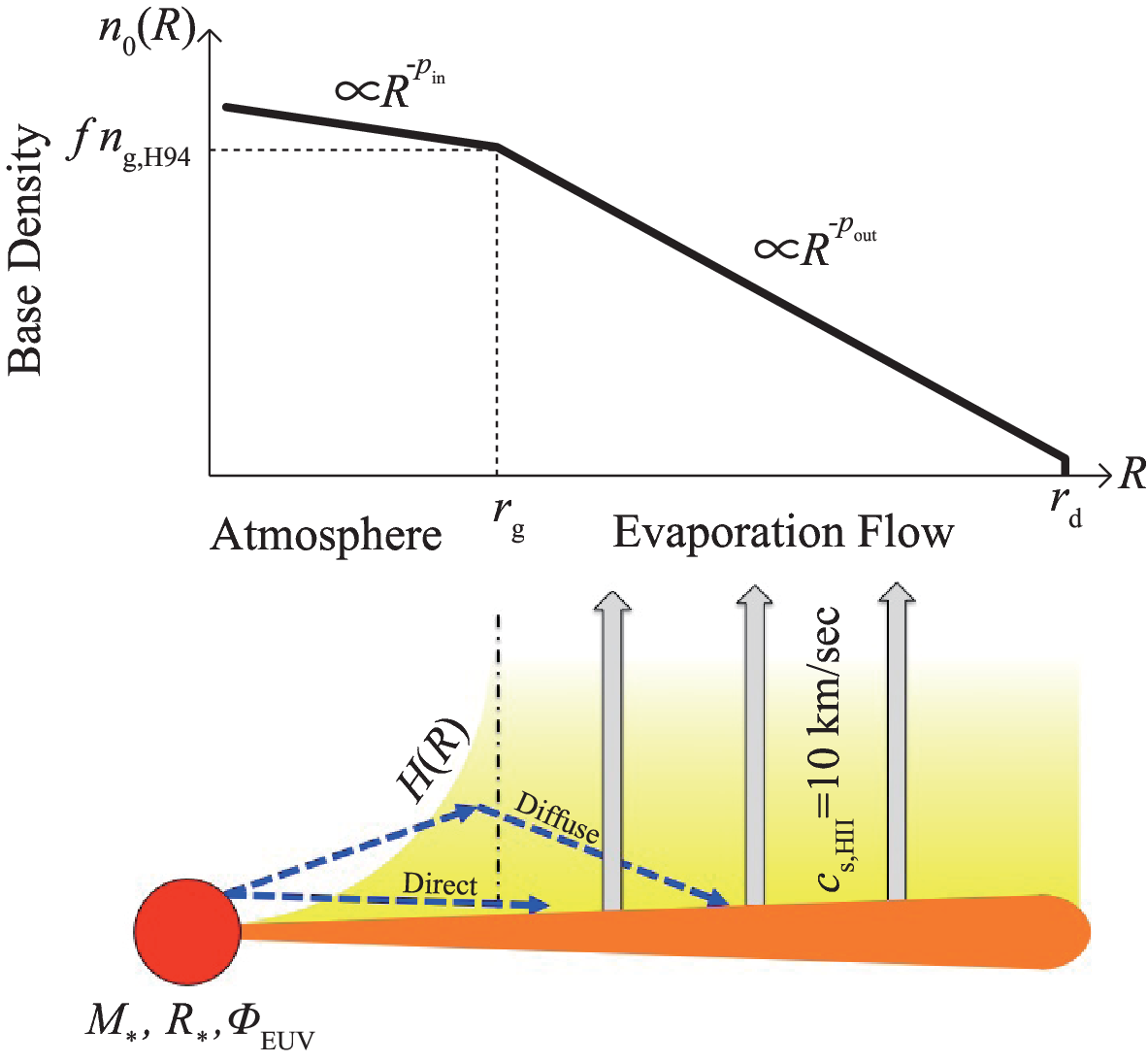}
\caption{Schematic view of our model. 
{\it Bottom:} Direct stellar radiation and diffuse radiation from recombination irradiate and ionize the disk surface, forming the atmosphere ($R<r_{\rm g}$) and the evaporation flow ($R>r_{\rm g}$). We assume a zero-thickness neutral disk, although the disk in this schematic view has finite thickness for illustration purposes. 
{\it Top:} Piecewise power-law distribution for 
the number density at the bottom of the H {\scriptsize \ II} region, or the base density 
(equations. \ref{eq_model1}-\ref{eq_model3}). }
\label{fig_ex}
\end{figure}

Photoevaporation is a mass-loss process from 
a circumstellar disk due to radiative heating 
either by its central star or by external objects 
\citep{bal82}.
We evaluate the mass-loss rate that results from 
EUV heating by the central star.
Below, we first describe the basics of photoevaporation;
Figure \ref{fig_ex} presents a schematic view of our model.
%{\bf
We use the axisymmetric cylindrical coordinates $R$ for the distance from the symmetric axis and $Z$ for the height from the equatorial plane.
%}

\subsection{Photoevaporation}
We consider photoevaporation from the disk 
around a star with mass $M_*$.
Because the density above the disk is lower than that 
in the equatorial plane, EUV radiation
irradiates the upper surfaces of the disk, and
the irradiated thin surface of the disk is ionized. 
The ionized gas is heated to a temperature of $\sim10^4{\rm K}$, which increases the sound speed to $c_{\rm s,HII} \approx 10{\rm km \ s^{-1}}$,
and forms the H {\scriptsize \ II} region above the disk.
The characteristic scale length, the so-called gravitational radius, is defined as the radius in which the Keplerian velocity ($\simeq$ escape velocity) is equal to $c_{\rm s,HII}$,
\begin{eqnarray}
	r_{\rm g} = \frac{GM_{*}}{c_{\rm s,HII}^2} \simeq 70 \left( \frac{M_{*}}{10M_{\odot}} \right) {\rm AU}.
	\label{eq_rg}
\end{eqnarray}
Inside the gravitational radius, $r \lesssim r_{\rm g}$, the ionized gas is gravitationally bound and 
an ionized atmosphere is created above the un-ionized 
disk.
However, outside the gravitational radius, $r \gtrsim r_{\rm g}$, the thermal energy of the ionized gas is greater than its gravitational energy;
thus, the ionized gas is unbound.
Therefore, the ionized gas flows away from the disk surface, and the gas is photoevaporated.
Because the photoevaporation  flow is driven by thermal pressure, the flow velocity is approximately given 
by the sound speed $c_{\rm s,HII}$.
Although the definition of the gravitational radius 
slightly changes if hydrodynamic effects such as gas pressure, radiation pressure, and angular momentum
are included \citep{lif03,fon04,mck08}, we here use eq.(\ref{eq_rg}) as the gravitational radius for easier comparison 
with HJLS94.
Our results in Section \ref{sec_analytic}
later show that the photoevaporation rate does not 
depend on the gravitational radius. 
This justifies our approximation for $r_{\rm g}$.

Because the photoevaporation flow velocity is $\simeq c_{\rm s, HII}$, 
the mass-loss flux per unit area at radius $R$ is
\begin{eqnarray}
	\dot{\Sigma}_{\rm PE}(R) = m_{\rm H} c_{\rm s, HII} n_{0}(R),
	\label{eq_sigdotpr}
\end{eqnarray}
where $m_{\rm H}$ is the proton mass, and $n_0(R)$ is the number density at the bottom of the ionization layers,
which is the boundary of the ionized H {\scriptsize \ II} region and the unionized disk.
We hereafter refer to $n_0(R)$ the base density.
The evaporation flow launches from the disk outside 
$r_{\rm g}$ to form an annulus with $r_{\rm g} < R < r_{\rm d}$, where $r_{\rm d}$ is 
the outer radius of the disk.
By summing the contribution from both the upper and lower surfaces,
the total photoevaporation rate from the disk is
\begin{eqnarray}
	\dot{M}_{\rm PE}
	&=& 2 \int_{r_{\rm g}}^{r_{\rm d}} 2\pi R \dot{\Sigma}_{\rm PE}(R) dR
	\nonumber \\
	&=& 4\pi m_{\rm H} c_{\rm s, HII} \int_{r_{\rm g}}^{r_{\rm d}} n_{0}(R) R dR.
	\label{eq_Mdotpe}
\end{eqnarray}
For a given base density distribution $n_0(R)$, 
we can calculate the photoevaporation rate.
HJLS94 estimated the base density distribution 
from the RT calculation with the 1+1D approximation
and derived the photoevaporation rate in equation (\ref{eq_pe_rate_h94}).
In this study, we repeat a similar analysis by using 
the 2D RT calculation.

\subsection{Ionized gas structure}
Following HJLS94,
we assume that (1) the steady-state density distribution
and (2) the thin neutral gas disk.
From these assumptions, the ionization front is located at $Z=0$.
%{\bf
We confirmed that our results with this thin-disk approximation are consistent with 
the hydrodynamic simulation for a finite-thickness disk (Section 5.2).
%}
Although the exact temperature of the ionized gas 
slightly depends on detailed heating and cooling 
processes, we retained the typical value of $T_{\rm HII}=10^4{\rm K}$.

\subsubsection{Vertical structure}
Inside the gravitational radius ($R<r_{\rm g}$), 
the bound ionized gas forms an atmosphere above the disk.
The gas is vertically hydrostatic in this region,
\begin{eqnarray}
	n(R,Z) = n_0(R) \exp\left( -\frac{Z^2}{2H^2} \right) 
\end{eqnarray}
where the scale height is expressed as
\begin{eqnarray} 
	H(R) = \frac{c_{\rm s,HII}}{\Omega_{\rm K}}
	= r_{\rm g} \left( \frac{R}{r_{\rm g}} \right)^{3/2}.
	\label{eq_nin}
\end{eqnarray}
On the contrary, the ionized gas will evaporate 
away in an outer flow region ($R>r_{\rm g}$).
In this region, we assume that the density is vertically constant,
\begin{eqnarray}
	n(R,Z) = n_0(R).
	\label{eq_nout}
\end{eqnarray}
Although this assumption fails in the upper region of $Z>R$ as shown in the hydrodynamic simulation by \citet{fon04},
we apply the concept for simplicity and 
for a comparison with HJLS94.
In consideration of our result 
that the density distribution in the lower layers of $Z\ll R$ is more important, 
this assumption dose not significantly affect our conclusion.

\subsubsection{Radial structure}
%{\bf
The vertical structure of the ionized region, which includes the atmosphere and evaporation flow, is given by equations (\ref{eq_nin}) and (\ref{eq_nout}).
To specify the radial density distribution,
we need the base density distribution, $n_0(R)=n(R,Z=0)$.
Here we suppose that the disk is neutral and geometrically thin
and that the ionized gas pervades above the neutral disk.
Thus, for a plausible density profile, the ionization front must coincide with the disk surface at $Z=0$.
With densities higher than the plausible value, a neutral zone appears above the disk.
This situation is inconsistent, because the vertical structure at $Z>0$ is only valid for hot ionized gas (equations \ref{eq_nin} and \ref{eq_nout}).
On the contrary, if the density is very low,
ionizing photons can reach the disk and ionize its upper layer,
which also conflicts with our assumption of the neutral gas disk.
In the following equation, by calculating the ionization structure for various base-density profiles,
we search for the plausible density profile, which places the ionization front at $Z=0$.
%}

In searching for the plausible base-density profile, 
we assume the piecewise power-law distribution with 
different exponent indices inside or outside the gravitational radius as in HJLS94:
\begin{eqnarray}
	n_0(R) = f n_{\rm g,H94} \left( \frac{R}{r_{\rm g}} \right)^{-p}, \label{eq_model1}
\end{eqnarray}
where
\begin{eqnarray}
	n_{\rm g, H94}
	&=& 1.8 \times 10^7 \left( \frac{\Phi_{\rm EUV}}{10^{49} {\rm s^{-1}} } \right)^{1/2}
	\nonumber \\
	&&~~~\times \left( \frac{r_{\rm g}}{10^{15}{\rm cm}} \right)^{-3/2} {\rm cm^{-3}}, \label{eq_model2} \\ 
	p &=& \left\{
	\begin{array}{ll}
		p_{\rm in} \ \ &{\rm for} \ R<r_{\rm g},\\
		p_{\rm out} \ \ &{\rm for} \ R>r_{\rm g},
	\end{array}
	\right. \label{eq_model3}
\end{eqnarray}
where $f,\ p_{\rm in}$, and $p_{\rm out}$ are dimensionless parameters. 
With the parameters set to $f=0.9,\ p_{\rm in}=1.5$, and $p_{\rm out}=2.5$, our base density reproduces that of HJLS04 model.
The normalization density $n_{\rm g, H94}\tbond C_{\rm H94} (3\Phi_{\rm EUV}/4\pi \alpha_{\rm B} r_{\rm g}^3)^{1/2}$ is that of HJLS94 derived analytically, where $\alpha_{\rm B}
 $ is the recombination coefficient to excited states
 (so-called case B) and $C_{\rm H94}\simeq 0.2$ is the correction factor used to reproduce their numerical results.
We search for the plausible base-density profile included the following process:
For a given density profile, we calculate the transfer of ionizing radiation.
If the gas at the disk surface ($Z=0$) was not ionized,  
we reduce the base density.
On the contrary, if the disk surface was well ionized, i.e., the ionizing photons have not been consumed before reaching that point, we elevate the base density.
In this manner, we obtain the plausible 
base-density profile iteratively.
In conclusion, we determine the parameters for the plausible base density to be $f=0.9$ and $p_{\rm in}= p_{\rm out}=1.5$, which differs from those in the HJLS94 model (Section \ref{sec_results}).
Because the outer density distribution ($p_{\rm out}=1.5$) is shallower than that in the HJLS94 model 
($p_{\rm out}=2.5$), 
the total mass-loss rate by photoevaporation depends on total area of the evaporating annulus
and thus the disk size (Section \ref{sec_analytic}), unlike in the case in HJLS94.

\subsection{Radiation transfer calculation}
We adopt the gray approximation for RT, 
where the frequency of ionizing photons is 
represented by the mean value. 
The equation of RT along a ray is
\begin{eqnarray}
	\frac{dI}{ds} = -(1-x) n\sigma_{\rm H}I 
+ \frac{\alpha_{1}x^2n^2}{4\pi} \epsilon,
	\label{eq_transfer}
\end{eqnarray}
where $x$ is the ionization degree, 
$\sigma_{\rm H}$ is the cross section of a 
hydrogen atom, $I$ is the irradiance intensity, 
$\alpha_1$ is the radiative recombination coefficient for the ground state, and $\epsilon$ is the mean energy of ionization photons.
The first term on the right-hand side represents the photon consumption by ionization, and the second term 
indicates the re-emission by recombinations directly to the ground state. 
Following HJLS94,
we neglect absorption and scattering by dust grains 
In the case of primordial star formation, as well as present-day star/planet 
formation, this may be an appropriate 
assumption if grains have significantly settled toward the equatorial plane.
However, the dust effects should be studied in future studies.

The ionization degree is obtained from the balance between photoionization and recombination,
\begin{eqnarray}
	\frac{4\pi x_{\rm HI}n\sigma_{\rm H}J}{\epsilon} = \alpha_{\rm A}x_{\rm HII}^2n^2,
	\label{eq_xeq}
\end{eqnarray}
where the mean intensity is
\begin{eqnarray}
	J = \frac{1}{4\pi}\int I d\Omega,
\end{eqnarray}
and $\alpha_{\rm A}$is the radiative recombination coefficient for all levels (so-called the case A).
Solving the RT equation (equation \ref{eq_transfer}) 
with the photoionization equilibrium (equation \ref{eq_xeq})
for a given base density distribution, 
we search for the parameter set $(f,p_{\rm in},p_{\rm out})$ 
for the plausible distribution as described in Section 2.2.2.

\subsection{Numerical settings}
We conduct axisymmetric two-dimensional RT 
calculations in the range of 
stellar mass $M_*=10$ -- $100M_*$ and 
EUV emissivity $\Phi_{\rm EUV} = 10^{49}$ -- $10^{51}{\rm s}^{-1}$.
The computational domain 
is a cylindrical region with 
radial and vertical coordinates $R<R_{\rm max}$ 
and $Z <Z_{\rm max}$, where $R_{\rm max}=Z_{\rm max} =r_{\rm g}$ and $10r_{\rm g}$. 
%{\bf
We assign an EUV source representing the central star at $(R,Z)=(0,0)$.
Although the source radius $R_*$ in our calculation ($> 0.2{\rm AU}$)
is greater than the actual stellar radius ($\la 0.1{\rm AU}$), 
we verify that the results are unchanged with varying
$R_*$ in the range from $0.02~R_{\rm max}$ to $0.003~R_{\rm max}$.
In the following section, we show the results with the highest resolution of $R_*=0.003R_{\rm max}$
($0.2{\rm AU}$ for $M_*=10M_\odot$ and $R_{\rm max}=r_{\rm g}$).
%}
Because finer structures are present in the inner region,
the radial and vertical grids are set to be spaced logarithmically. 
The EUV source radius is resolved with 10 grids 
and the entire computational domain by 70 grids.
We solve the transfer equation by the short characteristic method along rays that are tangential to 
cylinders with $R={\rm const.}$ \citep[Section 4,][]{sto92}.
The resolution in the zenithal angle is $\Delta \theta = \pi/1800$.
Because we solve ray-tracing by tangential plane methods, the resolution of azimuthal angle changes with spatial position. 
We verify the accuracy of our 2D RT code by 
conducting the spherical Str\"{o}mgren test with the 
same resolution.

%%%%%%%%%%%%%%%%%%
\section{Results} \label{sec_results}
In this section, we present the 2D calculation results,
which indicate that the exponent for the plausible 
base density is the same inside and outside the gravitational radius, $p_{\rm in}=p_{\rm in}=1.5$.
While this value is in disagreement with the conventional value reported 
by HJLS94, it is consistent with the results of the radiative hydrodynamic simulation by \citet{hos11} (Section \ref{sec_hosokawa}).

\begin{figure*}[t]
\epsscale{1.5}
\plotone{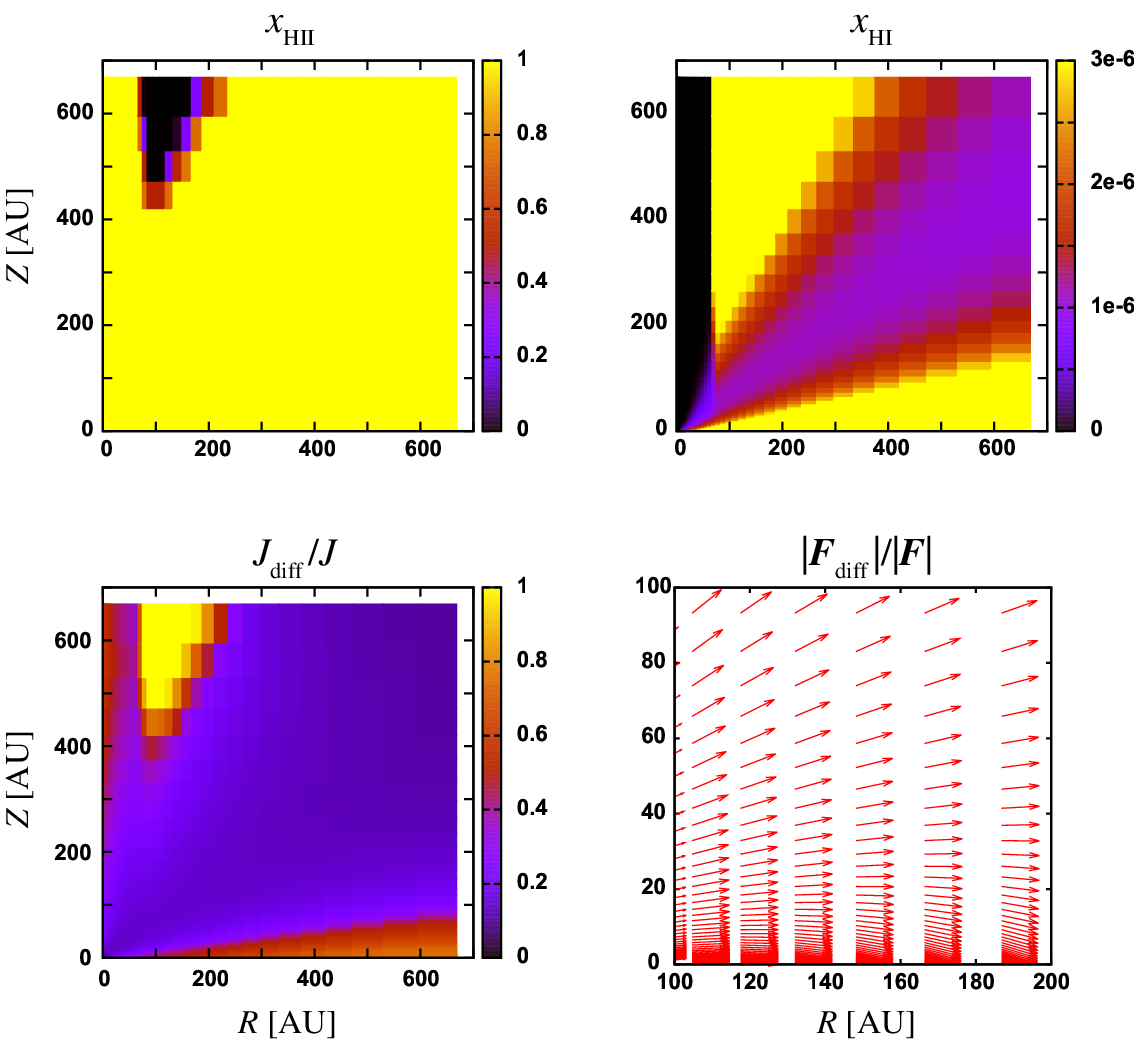}
\caption{
Spatial profiles of ionization degree $x_{\rm HII}$ ({\it top-left}),
neutral fraction $x_{\rm HI}$ ({\it top-right}),
diffuse radiation fraction $J_{\rm diff}/J$ ({\it bottom-left}),
and direction of diffuse flux ${\bm F}_{\rm diff}/|{\bm F}_{\rm diff}|$ ({\it bottom-right})
for the fiducial case with stellar parameters of $M_{*}=10M_\odot$ and $\Phi_{\rm EUV}=10^{49}{\rm sec}^{-1}$, base density of $f=0.9$ and $p_{\rm in}=p_{\rm out}=1.5$.
The gravitational radius is $r_{\rm g}\simeq 70 {\rm AU}$.
Note that the scale on the {\it bottom-right} panel is different so as to indicate the near base profile.
}
\label{fig_fiducial}
\end{figure*}

First, as the fiducial case, we show the results with stellar parameters
fixed at $M_{*}=10~M_\odot$ and $\Phi_{\rm EUV}=10^{49}~{\rm sec}^{-1}$.
From this result, we determine that the plausible base-density 
distribution is that with $f=0.9$ and 
$p_{\rm in}=p_{\rm out}=1.5$, for which the ionization front is located at the equatorial plane 
$Z=0$.
Figure \ref{fig_fiducial} shows the spatial profiles of 
ionization degree, neutral fraction, and diffuse radiation field 
in this case.
Inside the gravitational radius ($R<r_{\rm g}\simeq 70{\rm AU}$),
the density profile nearly agrees with that of the HJLS94 model 
($f=1, p_{\rm in}=1.5$).
However, outside the gravitational radius ($R>r_{\rm g}$), the density determined with our method is higher;
that is the disk surface ($Z=0$) can be ionized despite a density higher than that determined by HJLS94 model.

As assumed in HJLS94, 
the fraction of diffuse radiation $J_{\rm diff}/J$ 
is rather higher at the disk surface because of
re-emission in the atmosphere (Fig. \ref{fig_fiducial}, bottom-left).
However, this fraction reaches a maximum of $\sim0.5$;
thus, direct stellar radiation dominates in the entire region.
It should be noted the diffuse radiation flux, 
${\bm F}_{\rm diff}$, has a large radial component (Figure \ref{fig_fiducial}, bottom-right).
Such a radiation field cannot be properly expressed by 
1+1D treatment,
which emphasizes the importance of the 2D calculation.

\begin{figure}[t]
\epsscale{0.9}
%\begin{figure*}
%\epsscale{1.3}
\plotone{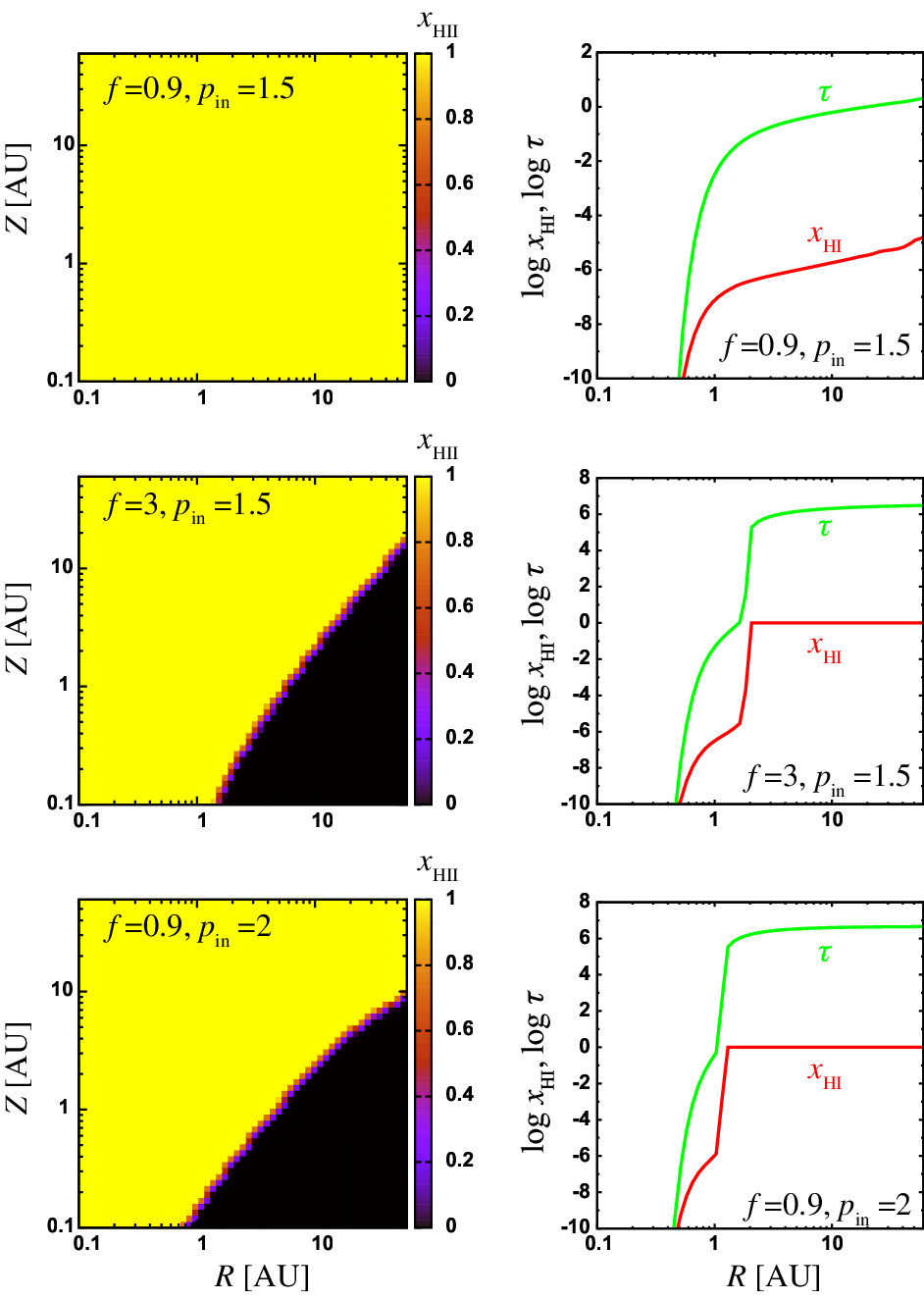}
\caption{Results with application of various $f$ and $p_{\rm in}$ showing ({\it left}) the ionization degree in the $R-Z$ plane and ({\it right}) the radial profiles of neutral fraction ($x_{\rm HI}$) and the optical depth from the central object ($\tau$) at the near base of $z=R_*$.
The density parameters are $f=0.9,~p_{\rm in}=1.5$ which is the plausible case, $f=3,~p_{\rm in}=1.5$, and $f=0.9,~p_{\rm in}=2$ from top to bottom, respectively. The ionization boundary always locates at $\tau\simeq1$.}
\label{fig_a}
%\end{figure*}
\end{figure}

Figure \ref{fig_a} shows the results for the most plausible density distribution and for other cases.
Here, to determine the dependence on the density normalization factor $f$ 
and the inner density exponent $p_{\rm in}$, the computational
domain is limited inside the gravitational radius, i.e., 
$R_{\rm max}=Z_{\rm max}=r_{\rm g}$.
For a density higher than the plausible value
($f>0.9\ {\rm or} \ p_{\rm in}>1.5$), 
the ionization front does not reach the disk surface at $Z=0$,
which is inconsistent with our assumption that the 
gas above the disk is ionized and has a high temperature.
These structures cannot be in the steady state, 
because the cold neutral gas would settle on the disk with  
the base density of ionized gas decreasing 
until the ionization front reaches the disk surface. 

Figure \ref{fig_a} shows that, in all the three cases, 
the ionization boundary, 
qt which the neutral fraction abruptly increases from $x_{\rm HI}\ll1$ to $\sim1$,
is located at the position where $\tau\sim1$ from the central star.
This fact indicates that direct stellar radiation is dominant in ionization,
and the gas is ionized as far as it reaches.
For the plausible distribution with 
($f=0.9$, $p_{\rm in}=p_{\rm out}=1.5$),
the optical depth remains $\sim1$ in a wide range of the radius,
which indicates that the ionization front is located $Z=0$ at all radii.

\begin{figure}[t]
\epsscale{.70}
\plotone{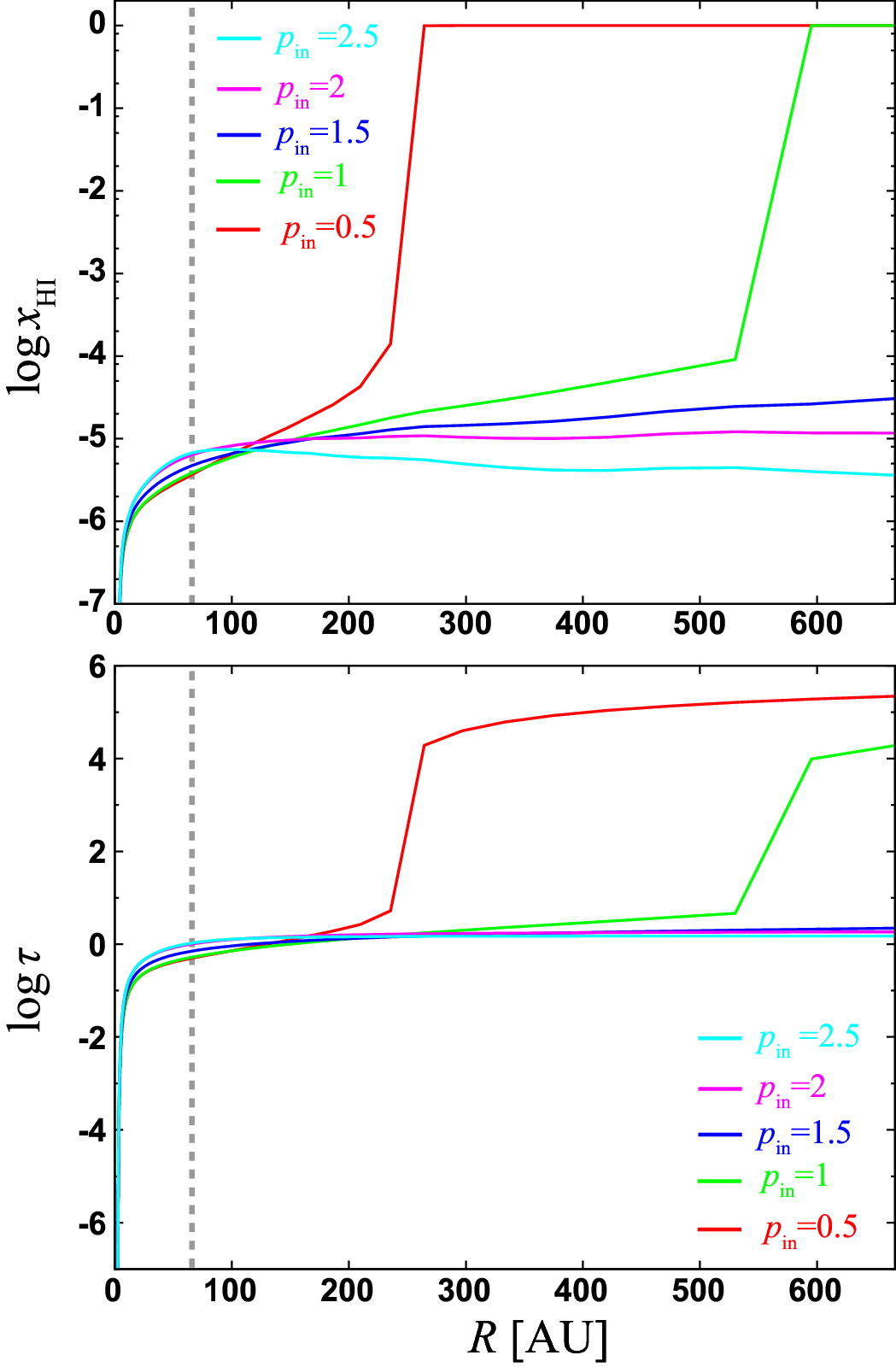}
\caption{Results with application of various $p_{\rm out}$ showing  ({\it top}) the radial profiles of neutral fraction and ({\it bottom}) the optical depth from the central object at the near base of $Z=R_*$. The dashed line indicates the gravitational radius, $R=r_{\rm g}$.}
\label{fig_d}
\end{figure}

To see the dependence on the outer density exponent $p_{\rm out}$, 
Fig. \ref{fig_d} shows the neutral fraction and the optical depth 
for the cases with five different values of $p_{\rm out}$ ($0.5, 1, 1.5, 2, $ and $2.5$), 
where the other parameters are fixed at ($f=0.9$, $p_{\rm in}=1.5$).
Inside the gravitational radius, both $x_{\rm HI}$ and 
$\tau$ behave similarly for various $p_{\rm out}$ values,
which indicates that the inner radiation field is not affected by 
the outer field.
For $p_{\rm out}<1.5$, the ionization front 
emerges above the disk when $\tau$ becomes $\simeq1$.  
For $p_{\rm out}>1.5$, the gas is completely ionized at the disk surface.
Thus, the case with $p_{\rm out}=1.5$ gives the plausible density
distribution, which produces the ionization front at the disk surface.

\begin{figure}[t]
\epsscale{.70}
\plotone{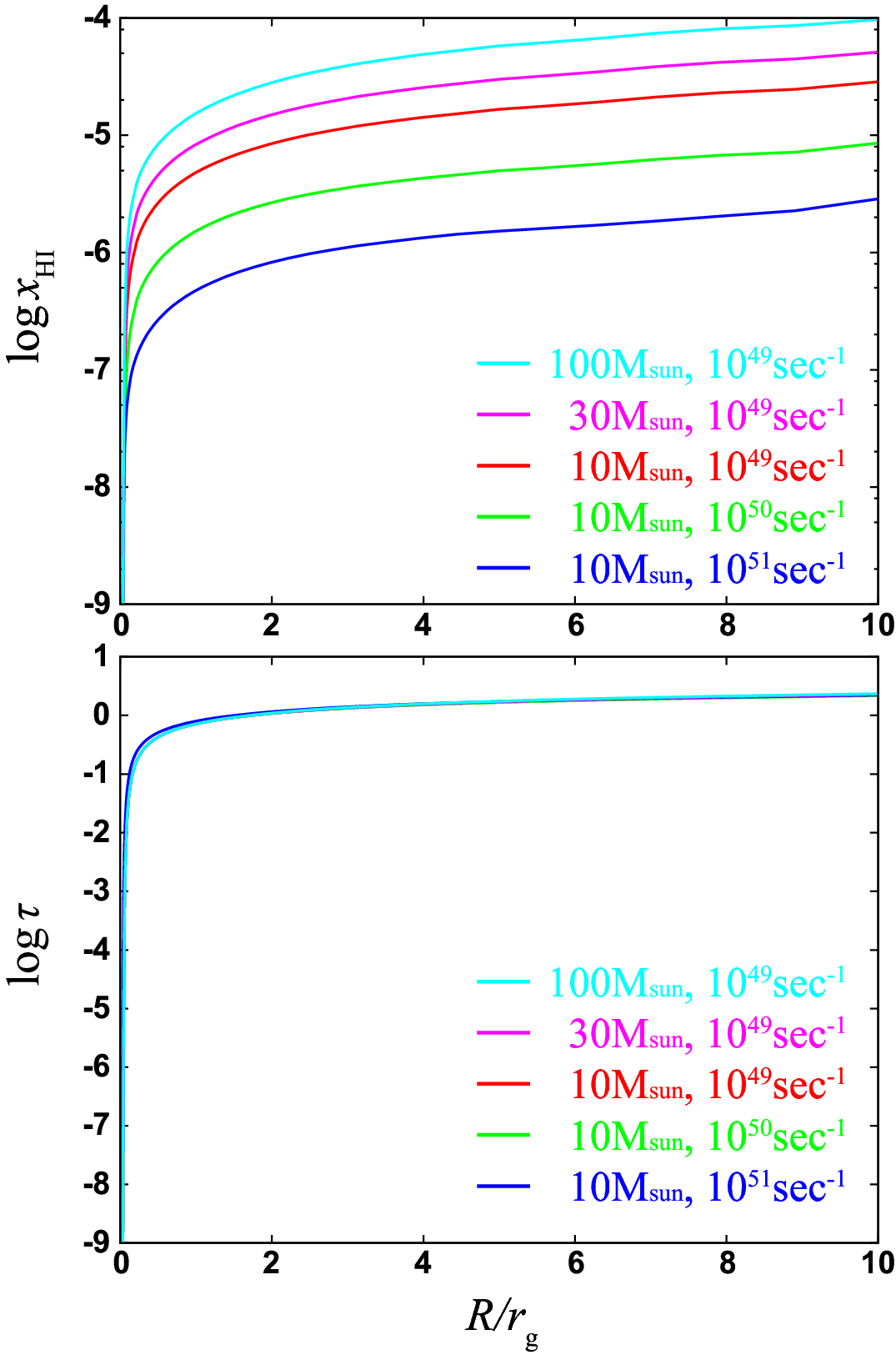}
\caption{Results with application of various $M_*$ and $\Phi_{\rm EUV}$
showing radial profiles of the neutral fraction and the optical depth from the central star.
In the bottom panel, all curves overlap.
Note that the horizontal axis is the radial distance normalized by the gravitational radius.}
\label{fig_e}
\end{figure}

Thus far, we have discussed cases with the stellar parameters 
$M_*=10M_\odot$ and $\Phi_{\rm EUV}=10^{49}{\rm sec}^{-1}$.
The same plausible density distribution is indicated
for other combinations of ($M_{\ast}$, $\Phi_{\rm EUV}$).  
Figure \ref{fig_e} shows the neutral fraction and optical depth
for the density distribution of $f=0.9$, $p_{\rm in}=p_{\rm out}=1.5$
for various stellar parameters ($M_*$, $\Phi_{\rm EUV}$).  
It should be noted that the horizontal axis is the radial distance 
normalized by the gravitational radius.
Although the neutral fraction depends on the stellar parameters, 
curves for the optical depth, which remains 
at $\simeq1$ for a wide range in radius, 
completely overlap.  
This result indicates that the density distribution of
$f=0.9$, $p_{\rm in}=p_{\rm out}=1.5$ is plausible
for any combination of ($M_*$, $\Phi_{\rm EUV}$).

Substituting $f=0.9$ and $p_{\rm in}=p_{\rm out}=1.5$ to eq.(\ref{eq_model3}), we obtain the base density distribution
\begin{eqnarray}
	n_0(R) = 1.6 \times 10^7 \left( \frac{\Phi_{\rm EUV}}{10^{49} {\rm s^{-1}} } \right)^{0.5}
	\left( \frac{R}{10^{15}{\rm cm}} \right)^{-1.5} {\rm cm^{-3}}.
	\label{eq_n_numerical}
\end{eqnarray}
This single power-law distribution is significantly differs from that of
the conventional HJLS94 model, where the density distribution 
follows the broken power-law with $p_{\rm in}=1.5$ and 
$p_{\rm out}=2.5$.
The stellar mass does not explicitly appear in the expression 
of equation (\ref{eq_n_numerical}),
and the base density depends on $M_*$ only through $\Phi_{\rm EUV}(M_*)$.

\section{Analytic expression for photoevaporation rate} \label{sec_analytic}
As demonstrated in the previous section,
stellar radiation dominates the diffuse radiation, and the ionization front always resides near $\tau\simeq1$.
On the basis of this fact, we analytically interpret the obtained base-density profile (equation \ref{eq_n_numerical}),
and we derive a new analytical expression for the photoevaporation rate for dust-free disks.

\begin{figure}[t]
\epsscale{1.}
\plotone{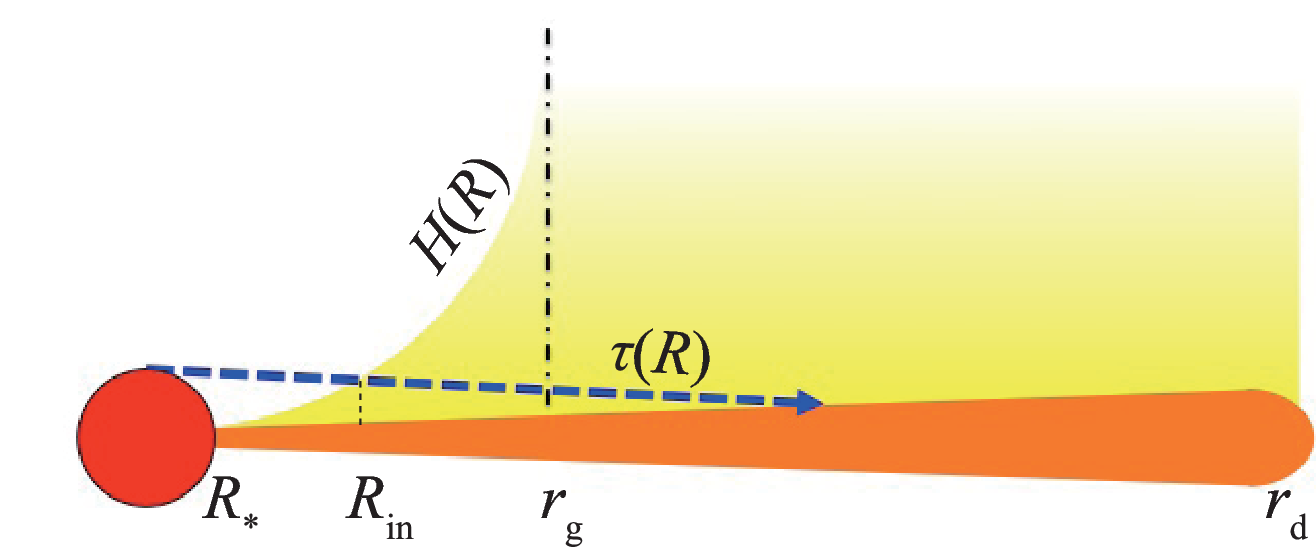}
\caption{Schematic view of the light path from the central star.}
\label{fig_ray}
\end{figure}

Let us consider a ray from a star irradiating onto the disk surface (see Fig. \ref{fig_ray}).
In the vicinity of the star, where 
\begin{eqnarray}
	R < R_{\rm in} &=& (r_{\rm g}R_*^2)^{1/3} \\
        &=& 0.3
	\left( \frac{M_*}{10M_\odot} \right)^{1/3}
	\left( \frac{R_*}{0.02{\rm AU}} \right)^{2/3} {\rm AU}, 
\end{eqnarray}
the scale height $H(R)$ of the ionized gas 
is lower than the height of the ray from the equatorial plane ($\sim R_*$),
and the density is low. 
The contribution of the gas inside $R_{\rm in}$ to the 
optical depth is negligible. 
Outside $R_{\rm in}$, the density along the ray is approximately 
given by the base density $n(R,Z)\simeq n_0(R)$,
because the ray travels below the scale height.
Then, the optical depth from the star can be written as 
\begin{eqnarray}
	\tau(R) = \int_{R_{\rm in}}^{R} x_{\rm HI}n_0(R') \sigma_{\rm H} dR'.
\end{eqnarray}
The neutral fraction $x_{\rm HII}(r,z=0)$ can be obtained  
from the photoionization equilibrium (equation \ref{eq_xeq}),
\begin{eqnarray}
	x_{\rm HI} \simeq  \frac{\alpha_{\rm A} \epsilon n}{4\pi \sigma_{\rm H} J},
	\label{eq_x}
\end{eqnarray}
where the case A value $\alpha_{\rm A}$ is used as the recombination coefficient,
because we consider only direct stellar radiation and ignore re-emitted diffuse 
radiation.
In this equation, 
we assume that the neutral fraction is small ($x_{\rm HI}\ll1$), 
and the optical depth along the ray is also small ($\tau\ll1$);
%$\tau\la1$ is used for the plausible density, as discussed in the previous section.
The direct stellar radiation intensity in $r\gg R_{*}$ is
\begin{eqnarray}
	J_* = \frac{\epsilon \Phi_{\rm EUV} }{16\pi^2r^2},
	\label{eq_Jst}
\end{eqnarray}
which is evaluated by the radiation from the half side of the star, which considers shielding by the optically thick disk.
From these values, we obtain the optical depth as,
\begin{eqnarray}
	\tau(R) \simeq \int_{R_{\rm in}}^{R}
	\frac{4\pi \alpha_{\rm A}  n_0^2 R'^2}{\Phi_{\rm EUV}} dR'.
	\label{eq_tau0}
\end{eqnarray}

Assuming a single power-law distribution for the 
base density, 
$n_0 \propto r^{-p}$, the optical depth is approximately evaluated as
\begin{eqnarray}
	\tau(R) \sim \frac{4\pi \alpha_{\rm A}  n_0^2 R^3}{\Phi_{\rm EUV}} \propto R^{3-2p}.
	\label{eq_tau_analytic}
\end{eqnarray}
For the plausible density distribution, 
the ionization front is located at $Z=0$ for all radii or $\tau \sim 1 \propto R^0$.
Thus, from equation (\ref{eq_tau_analytic}), we obtain the exponent $p=3/2$.
If $p < 3/2$, the optical depth $\tau$ increases with $r$, and it does not
satisfy the requisite $\tau \sim R^0$.
Strictly speaking, with $p={3/2}$, 
$\tau$ is an increasing function as $\log R$, 
however, it should be noted that
$\log r$ is a much flatter function of $r$ than $R^{3-2p}$ with $p < 3/2$.
In contrast,
if $p > 3/2$, the base density $n_0$ rapidly decreases with $R$,
and the optical depth $\tau$ is dominated by the density in
the inner region ($R \simeq R_{\rm in}$).
Then, the optical depth in this case would also become constant with respect to the radius, $\tau(R) \propto R^0$.
However, if the density is lower than the plausible value and $\tau < 1$,
the base density would increase because of the direct stellar radiation and ionization
and would approach the plausible value, which has the exponent
$p = 3/2$.
On the other hand,
if the density is higher than the plausible value and $\tau > 1$,
the base density would decrease because ionizing radiation does not reach that point.
Therefore, in any case for $p > 3/2$, the density would  approach the plausible value.

We can also obtain the absolute value of the base density
from $\tau \simeq 1$ with $p = 3/2$ as,
\begin{eqnarray}
	n_0(R) = C\left( \frac{\Phi_{\rm EUV}}{4\pi \alpha_{\rm A} R^3} \right)^{1/2}
	~~~~~~~~~~~~~~~~~~~~~~~~~~~~
	\nonumber \\
	= 1.6 \times 10^7 \left( \frac{\Phi_{\rm EUV}}{10^{49} {\rm s^{-1}} } \right)^{1/2}
	\left( \frac{R}{10^{15}{\rm cm}} \right)^{-3/2} {\rm cm^{-3}},
	\label{eq_n_new}
\end{eqnarray}
where $C\simeq 0.4$ is the correction factor used to match
our numerical result (equation \ref{eq_n_numerical}).
It is evident that the agreement between the analytical and numerical results is 
satisfactory.  

Finally, we evaluate the photoevaporation rate by using 
the obtained base density from equation (\ref{eq_n_new}).
From equation (\ref{eq_sigdotpr}), 
the mass-loss flux from a unit area is
\begin{eqnarray}
	\dot{\Sigma}_{\rm PE}(R) = 6.0 \times 10^{-13}
	\left (\frac{\Phi_{\rm EUV}}{10^{49}{\rm sec}^{-1}} \right)^{1/2}
	\nonumber \\
	~~~~~~~~~~ \times \left (\frac{R}{1000{\rm AU}} \right)^{-3/2}
	{\rm g \ \ cm^{-2} \ \ s^{-1}}.
\end{eqnarray}
From equation (\ref{eq_Mdotpe}), the total evaporation rate from the entire disk is 
\begin{eqnarray}
	\dot{M}_{\rm PE} = 5.4 \times 10^{-5}
	\left (\frac{\Phi_{\rm EUV}}{10^{49}{\rm sec}^{-1}} \right)^{1/2}
	\nonumber \\
	~~~~~~~~~~ \times \left (\frac{r_{\rm d}}{1000{\rm AU}} \right)^{1/2}
	M_\odot {\rm yr}^{-1}.
	\label{eq_pe_rate}
\end{eqnarray}
Here, we assume that the disk size is significantly greater
than the gravitational radius ($r_{\rm d} \gg r_{\rm g} $).
However, for very massive stars ($>20~M_\odot$),
the gravitational radius defined by equation (\ref{eq_rg}) can be 
as large as the disk size.
Even in this case, if we also consider effects such as 
radiation pressure and angular momentum in addition 
to gravity, the {\it effective} gravitational radius will
be reduced below the disk radius \citep{lif03,fon04,mck08}.
It should be noted that 
our evaporation rate (equation \ref{eq_pe_rate}) depends 
on the disk radius rather than the gravitational radius, 
unlike that given by HJLS94.
This discrepancy occurs, because in the HJLS94 model, the base density 
steeply decreases as $R^{-5/2}$ outside the gravitational radius, 
whereas for the flat base-density distribution $R^{-3/2}$ 
in our model, annuli near the outer radius $r_{\rm d}$ 
dominate the evaporation rate.
Formally, equation (\ref{eq_pe_rate})  
coincides with the results of HJLS94 if we change $r_{\rm d}$ to 
$r_{\rm g}$.
Although the terms appear to differ in the expression, 
their numerical difference, $\sqrt{r_{\rm d}/r_{\rm g}}$, generally remains within an order of magnitude. 

\section{Discussion}
Photoevaporation has an important effect on 
the formations of stars and planets.
In this section, we discuss the impact of EUV photoevaporation 
on primordial 
star formation as well as present-day star/planetary formation 
by considering the new photoevaporation rate.
In addition, we describe the validity and the limitations of our model.

\subsection{Impacts of photoevaporation in star/planet formation}
\subsubsection{Primordial star formation}
Recent studies have demonstrated that the protostellar accretion 
of primordial stars is terminated by
EUV photoevaporation  
\citep{mck08,hos11,sta12, hos12b}.
Once the stellar mass exceeds $\sim 10M_\odot$, 
the stellar surface temperature becomes sufficiently high to emit 
a copious amount of EUV photons.
As a result, the conical H {\scriptsize \ II} regions begin 
to expand in the polar directions 
in a process known as the H {\scriptsize \ II} region breakout,
which significantly decreases the infall rate from the envelope to the disk.
Disk photoevaporation begins at that moment and increases with the growth of the stellar mass.
Finally, accretion ceases when the growing 
evaporation rate reaches the infall rate such that
 $\dot{M}_{\rm PE}=\dot{M}_{\rm infall}$.
%This process is  \citep[the schematic is shown in Fig. 1 of][]{mck08}.

\begin{figure}[t]
\epsscale{0.9}
\plotone{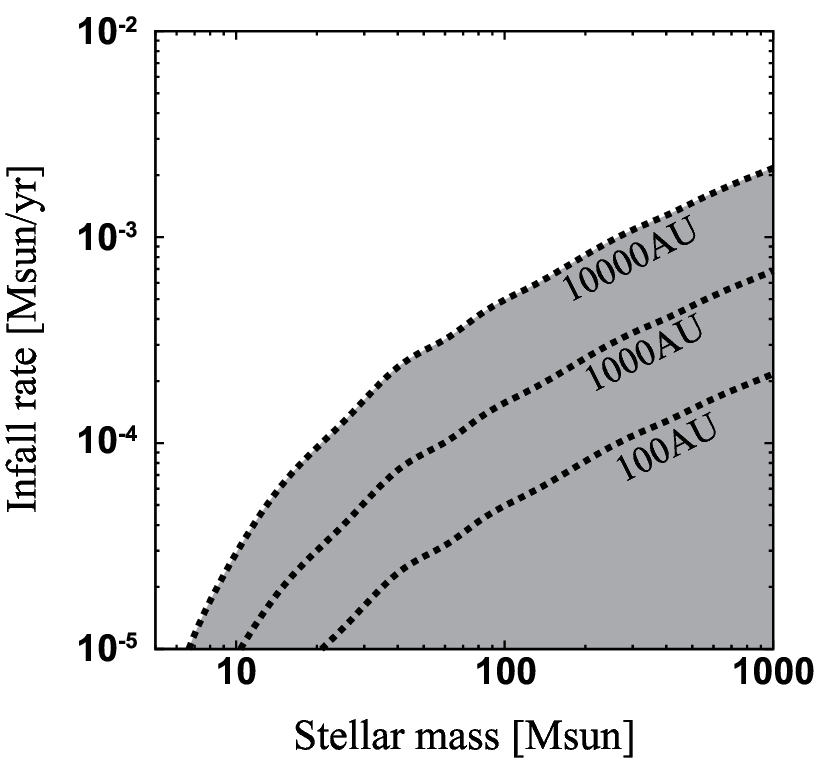}
\caption{Final mass of a primordial star obtained by equating the infall rate $\dot{M}_{\rm infall}$ with the photoevaporation rate $\dot{M}_{\rm PE}$.
The three curves represent cases in which the disk radius $r_{\rm d} = 100, 1000$, and $10000~ {\rm AU}$.
Formation in the lower-right region is prohibited by photoevaporation.
For example, in the case of $\dot{M}_{\rm infall}=10^{-4}M_\odot {\rm yr}^{-1}$ and $r_{\rm d}=1000{\rm AU}$, the final stellar mass is approximately $50M_\odot$.
We use the EUV emissivity $\Phi_{\rm EUV}$ of the zero-age main sequence star reported by \citet{sch02}.}
\label{fig_mass_limit}
\end{figure}

%{\bf
In Fig.\ref{fig_mass_limit},
we show the mass of primordial stars at the termination of accretion,
$\dot{M}_{\rm PE}=\dot{M}_{\rm infall}$,
estimated by our photoevaporation rate
for three disk sizes of $r_{\rm d}=100, 1000$, and $10000{\rm AU}$
(equation \ref{eq_pe_rate}).
The disk radius is $\sim100 - 10000 {\rm AU}$ for stars with mass of $10 $ -- $ 1000~M_\odot$,
as determined by the analytical model for primordial star formation developed by \citet{tan04}.
%}
Here we evaluate EUV emissivity
by assuming that the star is in the zero-age main sequence (ZAMS) phase \citep{sch02}.
EUV emissivity for ZAMS stars can be approximated as 
$\Phi_{\rm EUV}\simeq 1.26 \times 10^{47} (M_{\ast}/M_{\odot})^{1.4}~{\rm sec}^{-1}$ in the range $40M_\odot<M_{\ast}<1000M_\odot$.
Then, we obtain the stellar mass whereby the photoevaporation rate balances the accretion rate, 
\begin{eqnarray}
	M_{\ast} \simeq 55
	\left( \frac{ r_{\rm d}}{1000{\rm AU}} \right)^{-0.7}
	\left( \frac{ \dot{M}_{\rm infall} }{10^{-4}M_\odot{\rm yr}^{-1}} \right) ^{1.4}  M_{\odot}.
	\label{eq_finalmass}
\end{eqnarray}
Here we use the typical accretion rate of
$10^{-4}M_\odot{\rm yr}^{-1}$, which is smaller than the 
conventional rate of $\sim 10^{-3}M_\odot{\rm yr}^{-1}$ in primordial star formation without stellar feedback \citep{omu98,abe02,bro04,yos06},
because the accretion rate is reduced by approximately one order of magnitude before photoevaporation finally quenches the mass supply to the star owing to the H {\scriptsize \ II} region breakout \citep{mck08, hos11}.
The obtained mass of $55~M_\odot$ agrees well with the results of hydrodynamic simulations by \citet{hos11}, which indicate that the reduction of the infall rate through the H {\scriptsize \ II} region breakout,
along with photoevaporation, is essential in setting the final stellar mass.
It should be noted the final mass is smaller for the larger disk radius,
because the evaporation rate is proportional to $r_{\rm d}^{1/2}$.
Therefore, stars formed in pre-stellar cores with a larger angular 
momentum would be smaller because of the greater radius of the 
protostellar disk, in addition to the possible reduction of 
the infall rate due to centrifugal force.

The following caveat is to be noted regarding our adoption of the ZAMS EUV emissivity: 
Although stars generally reach the ZAMS by the time of accretion termination, 
this is rather a result of reduced accretion, i.e., longer accretion time than the stellar Kelvin-Helmholtz time,
through feedback during the preceding phase. 
In the pre-ZAMS phase, the stellar structure and 
the accretion rate 
interact in a manner such that a more rapid infall 
results in a larger stellar radius.
Thus, EUV emissivity is smaller,
which in turn, results in weaker feedback to the infall
\citep{hos11,hos12b}.
To treat the feedback in the pre-ZAMS phase,
the elaborate modeling of stellar evolution is necessary.

\subsubsection{Present-day star/planet formation}
Before discussing present-day star/planet formation,
it should be noted that
our derived EUV photoevaporation rate
ignores the scattering and absorption by dust grains,
similar to the cases in the conventional HJLS94 model.
\citet{ric97} conducted hydrodynamical simulation to demonstrate that in some limited cases,
dust scattering process increases the photoevaporation rate by a factor of $\sim 2$.
We thus regard our photoevaporation rate as  
a rough value with a factor of a few degrees of uncertainty.
Therefore, comprehensive research about the dust effect in this process is necessary.

In the formation of present-day massive stars,
the maximum stellar mass may be determined by the combination of photoevaporation and radiation pressure.
Precious studies have demonstrated that the longstanding issue of a radiation pressure barrier in massive star formation can be overcome by the shielding property of disk accretion
%{\bf
\citep{nak89, jij96, kru09, kui10, tan11}.
%}
However,
the radiation pressure still has a strong effect in depleting the infall rate from the dusty envelope \citep{kui12}.
With the depletion of the infall rate by radiation pressure, photoevaporation can terminate mass accretion onto stars.
Therefore,
even with an observationally claimed high accretion rate of
$10^{-4} $ -- $10^{-3}{M_\odot {\rm yr}^{-1}}$, which is 
similar to that in the first star formation,
the final masses of these stars are expected to be smaller than those of the first stars.

EUV photoevaporation has less importance
in the dissipation of protoplanetary disks around low-mass stars.
Because such stars emit few EUV photons, 
an increased amount of transmission radiation,
such as FUV radiation and/or 
X-rays, is considered to dominate disk photoevaporation 
\citep{gor09, gor09b, owe10,owe12}.
In fact, although the EUV evaporation rate determined by our model is higher than that by the HJLS94 model,
the value falls below the rate determined by FUV and X-rays by more than one order of magnitude 
for disks around low-mass stars.
%{\bf
For the dissipation of protoplanetary disks,
the role of UV radiation from nearby massive stars should be also be considered,
because most stars are born as members of star clusters
\citep{joh98, ada04, fat08, hol11,tho13}.
%}

\begin{figure*}[t]
\epsscale{1.5}
%\begin{figure*}
%\epsscale{2.0}
\plotone{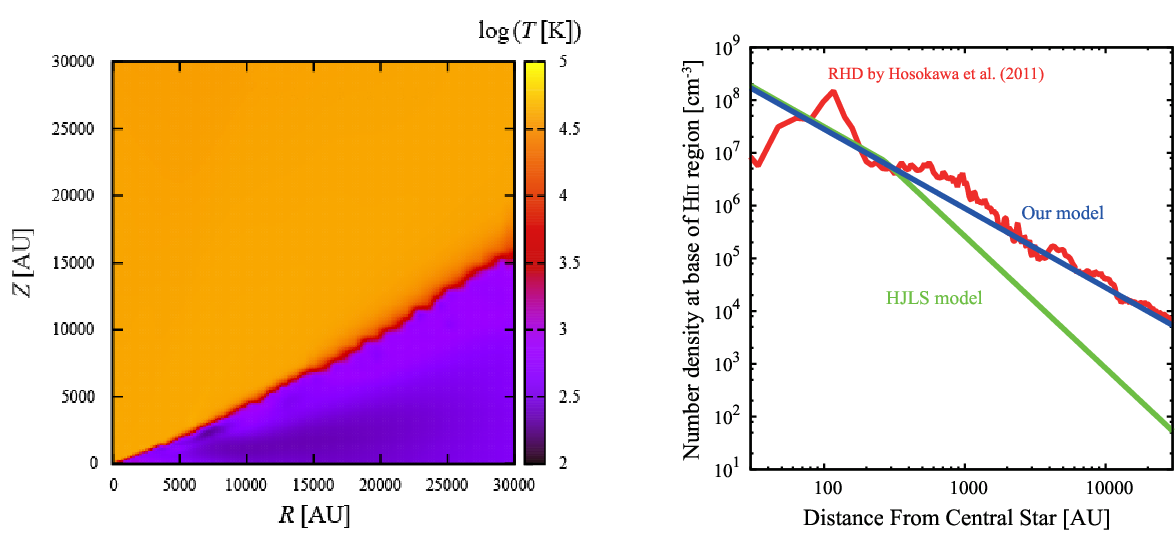}
\caption{{\it (left):} Temperature distribution at the photoevaporation stage when $M_*\simeq40~M_\odot$ and $\Phi_{\rm EUV}\simeq10^{50}~{\rm sec}^{-1}$, from the hydrodynamical simulation from \citet{hos11}.
We can see the neutral disk at the equatorial plane ($T<10^4{\rm K}$) and the ionized region above the disk ($T>10^4{\rm K}$).
{\it (right):} Comparison of our density models, that of HJLS94, and the numerical result by \citet{hos11}
are indicated by blue, green, and red lines, respectively.
The results of our model are consistent with the numerical results.}
\label{fig_hosokawa}
%\end{figure*}
\end{figure*}

\subsection{Comparison with numerical simulations} \label{sec_hosokawa}
In this section, we discuss the validity of our model by 
comparing our steady density model with the numerical results reported by \citet{hos11},
who investigated photoionization feedback in primordial star formation
by combining two-dimensional radiative hydrodynamics for the envelope 
with the stellar evolution calculation.
In their simulation, direct stellar radiation is solved by ray-tracing, while diffuse radiation is treated with the flux limited diffusion approximation.
The left-hand panel of Fig. \ref{fig_hosokawa} shows the temperature structure of the photoevaporation disk at the stellar mass 
$M_*\simeq40M_\odot$ with $\Phi_{\rm EUV}\simeq10^{50}{\rm sec}^{-1}$, as determined by their fiducial model.
The disk is in fact in the quasi-steady state in the photoevaporation epoch.
The neutral disk or troid ($\lesssim 10^4{\rm K}$) extends on the equatorial plane and ionized gas ($\gtrsim 10^4{\rm K}$) 
occupies the region above the disk.
In their simulation, accretion is terminated 
by the photoevaporation at a final stellar mass of $43M_\odot$.

In the right-hand panel of the figure,
the density at the ionization front from the numerical simulation 
of \citet{hos11} and our determined base density in the case of 
$p_{\rm in}=p_{\rm out}=1.5$ are presented as 
a function of the radius.
In addition, the base density determined by the HJLS94 model 
($p_{\rm in}=1.5, p_{\rm out}=2.5$) is shown.
A comparison of these values clearly reveals that our results, 
including the exponent of $1.5$ and also the absolute value, 
agree with the numerical results.
Thus, we conclude that our steady assumption is 
valid even when realistic hydrodynamical effects are included.
It should be noted that our result derived with the thin-disk approximation 
agrees well with that of the simulation, in which the disk has finite thickness.
Therefore, our model is applicable as long as direct stellar radiation irradiates 
the disk surface in a wide range of the radius.
Our model is also consistent with
results by \citet{hos11} such that direct stellar 
radiation dominates photoionization.

\subsection{Limitation of our model} \label{sec_limitation}
It should be noted that our determined photoevaporation rate is only 
a rough approximation and carries some uncertainty.
In our model, for example, we assume that the ionized-gas 
temperature is constant at $T_{\rm HII}=10^4{\rm K}$
and that the flow velocity is given by the sound speed at
$v_{\rm PE}\sim c_{\rm s, HII} \propto T_{\rm HII}^{1/2}$.
Both of these assumptions require modification in some circumstances.
For the former, the ionized gas near primordial stars 
can exhibit higher temperatures owing to higher stellar surface temperatures and less efficient cooling
\citep[$\sim4\times10^4{\rm K}$ in][]{hos11}.
For the latter, according to hydrodynamical simulation conducted by \citet{fon04}, the flows at the ionization boundary 
are slightly slower than the sound speed, which makes 
the mass-loss rate smaller by a factor of two.
For a more accurate evaluation, 
sophisticated radiative hydrodynamical modeling is required.

In this study, we ignore the effects of dust grains.
For solar metallicity, the scattering of radiation by 
dust in the disk atmosphere enhances the irradiation EUV flux
onto the disk surface and increases 
the photoevaporation rate by a factor of approximately two, 
according to radiation hydrodynamical simulation by conducted
\citet{ric97}.
However, their calculation was limited to several combinations of 
stellar mass and disk size with the approximated RT calculation.
Further calculations employing wide ranges of these parameters are needed for discussing their influence.
The dependence on the disk radius is particularly interesting,
because in the simulation conducted by \citet{ric97}, 
the mass loss from the outer region 
dominated the total evaporation rate.

In this study, we illustrate the case of single star formation.
However,
binaries or small multiples are also expected to be formed
both in cases of first star formation \citep{mac08,sta10,cla11}
and of present-day massive star formation \citep{kra06,kru09,pet10}.
Because the infalling material is divided into multiple stars,
the individual stars would be smaller than that 
in the single star case.
This process is known as fragmentation-induced starvation \citep{pet10}.
Therefore, the final mass estimated by equation (\ref{eq_finalmass}) may be the upper limit;
further investigation is necessary for accurate estimation.

%%%%%%%%%%%%%%%%%%
\section{Summary}
The photoevaporation of circumstellar disks by extreme 
ultraviolet radiation plays an important role 
in star formation in the present-day as well as early universe.
In this study, we revisit the limitation present in the conventional 1+1D approximation model developed by Hollenbach et al. (1994)
in dust-free case by introducing an updated axisymmetric 2D model.
Unlike that in the conventional model, the density distribution 
at the photoionization front located just above the disk,
known as the base density' distribution,
is represented by a single-exponent power law with an index of $-3/2$.
In our model, the total photoevaporation rate depends on the outer disk 
radius (equation \ref{eq_pe_rate}) in contrast to that in the conventional 
model, which depends on the gravitational radius.
Although we have derived this base-density distribution under the 
steady-state assumption, we have confirmed that
the results of our model are consistent with those of the radiative hydrodynamical simulation 
conducted by \citet{hos11}.

%%%%%%%%%%%%%%%%%%
\acknowledgments
We are grateful to Takashi Hosokawa, Jonathan C. Tan, and Taku Takeuchi for fruitful discussions.
This study is supported in part by the Grants-in-Aid by the Ministry of Education, Science and Culture of Japan 
(2157031:KT,  21540434, 24340102:TN,  2124402, 21684007:KO).
This study is also supported by Global COE program ``From the Earth to Earths.h

%\appendix




\clearpage


\begin{thebibliography}{}

\bibitem[Abel et al.(2002)]{abe02}
	Abel, T., Bryan, G.~L., \& Norman, M.~L.\ 2002, Science, 295, 93 

\bibitem[Adams et al.(2004)]{ada04}
	Adams, F.~C., Hollenbach, D., Laughlin, G., \& Gorti, U.\ 2004, \apj, 611, 360 

\bibitem[Armitage(2011)]{arm11}
	Armitage, P.~J.\ 2011, \araa, 49, 195 

\bibitem[Bally \& Scoville(1982)]{bal82}
	Bally, J., \& Scoville, N.~Z.\ 1982, \apj, 255, 497 

\bibitem[Bromm et al.(1999)]{bro99}
	Bromm, V., Coppi, P.~S., \& Larson, R.~B.\ 1999, \apjl, 527, L5 

\bibitem[Bromm \& Loeb(2004)]{bro04}
	Bromm, V., \& Loeb, A.\ 2004, \na, 9, 353 

\bibitem[Clarke et al.(2001)]{cla01}
	Clarke, C.~J., Gendrin, A., \& Sotomayor, M.\ 2001, \mnras, 328, 485 

\bibitem[Clark et al.(2011)]{cla11}
	Clark, P.~C., Glover, S.~C.~O., Smith, R.~J., et al.\ 2011, Science, 331, 1040 

\bibitem[Fatuzzo \& Adams(2008)]{fat08}
	Fatuzzo, M., \& Adams, F.~C.\ 2008, \apj, 675, 1361

\bibitem[Font et al.(2004)]{fon04}
	Font, A.~S., McCarthy, I.~G., Johnstone, D., \& Ballantyne, D.~R.\ 2004, \apj, 607, 890 

\bibitem[Gorti \& Hollenbach(2009)]{gor09}
	Gorti, U., \& Hollenbach, D.\ 2009, \apj, 690, 1539 
	
\bibitem[Gorti et al.(2009)]{gor09b}
	Gorti, U., Dullemond, C.~P., \& Hollenbach, D.\ 2009, \apj, 705, 1237 


\bibitem[Holden et al.(2011)]{hol11}
	Holden, L., Landis, E., Spitzig, J., \& Adams, F.~C.\ 2011, \pasp, 123, 14 

\bibitem[Hollenbach et al.(1994)]{hol94}
	Hollenbach, D., Johnstone, D., Lizano, S., \& Shu, F.\ 1994, \apj, 428, 654 (HJLS94)

\bibitem[Hosokawa et al.(2011)]{hos11}
	Hosokawa, T., Omukai, K., Yoshida, N., \& Yorke, H.~W.\ 2011, Science, 334, 1250

\bibitem[Hosokawa et al.(2012)]{hos12b}
	Hosokawa, T., Yoshida, N., Omukai, K., \& Yorke, H.~W.\ 2012, \apjl, 760, L37 

%\bibitem[Hosokawa et al.(2012)]{hos12b}
%	Hosokawa, T., Yoshida, N., Omukai, K., \& Yorke, H.~W.\ 2012b, arXiv:1210.3035 

\bibitem[Jijina \& Adams(1996)]{jij96}
	Jijina, J., \& Adams, F.~C.\ 1996, \apj, 462, 874

\bibitem[Johnstone et al.(1998)]{joh98}
	Johnstone, D., Hollenbach, D., \& Bally, J.\ 1998, \apj, 499, 758 

\bibitem[Kratter \& Matzner(2006)]{kra06}
	Kratter, K.~M., \& Matzner, C.~D.\ 2006, \mnras, 373, 1563 

\bibitem[Kraus et al.(2010)]{kra10}
	Kraus, S., Hofmann, K.-H., Menten, K.~M., et al.\ 2010, \nat, 466, 339 

\bibitem[Krumholz et al.(2009)]{kru09}
	Krumholz, M.~R., Klein, R.~I., McKee, C.~F., Offner, S.~S.~R., \& Cunningham, A.~J.\ 2009, Science, 323, 754 

\bibitem[Kuiper et al.(2010)]{kui10}
	Kuiper, R., Klahr, H., Beuther, H., \& Henning, T.\ 2010, \apj, 722, 1556 

\bibitem[Kuiper et al.(2012)]{kui12}
	Kuiper, R., Klahr, H., Beuther, H., \& Henning, T.\ 2012, \aap, 537, A122 

\bibitem[Liffman(2003)]{lif03}
	Liffman, K.\ 2003, \pasa, 20, 337 

\bibitem[Machida et al.(2008)]{mac08}
	Machida, M.~N., Omukai, K., Matsumoto, T., \& Inutsuka, S.-i.\ 2008, \apj, 677, 813 

\bibitem[McKee \& Tan(2008)]{mck08}
	McKee, C.~F., \& Tan, J.~C.\ 2008, \apj, 681, 771 

\bibitem[Nakano(1989)]{nak89}
	Nakano, T.\ 1989, \apj, 345, 464 

\bibitem[Omukai \& Nishi(1998)]{omu98}
	Omukai, K., \& Nishi, R.\ 1998, \apj, 508, 141

\bibitem[Omukai \& Palla(2003)]{omu03}
	Omukai, K., \& Palla, F.\ 2003, \apj, 589, 677 

\bibitem[Owen et al.(2010)]{owe10}
	Owen, J.~E., Ercolano, B., Clarke, C.~J., \& Alexander, R.~D.\ 2010, \mnras, 401, 1415 

\bibitem[Owen et al.(2012)]{owe12}
	Owen, J.~E., Clarke, C.~J., \& Ercolano, B.\ 2012, \mnras, 422, 1880 

\bibitem[Peters et al.(2010)]{pet10}
	Peters, T., Klessen, R.~S., Mac Low, M.-M., \& Banerjee, R.\ 2010, \apj, 725, 134 
\bibitem[Richling \& Yorke(1997)]{ric97}
	Richling, S., \& Yorke, H.~W.\ 1997, \aap, 327, 317 
	
\bibitem[Schaerer(2002)]{sch02}
	Schaerer, D.\ 2002, \aap, 382, 28 

\bibitem[Shu et al.(1993)]{shu93}
	Shu, F.~H., Johnstone, D., \& Hollenbach, D.\ 1993, \icarus, 106, 92 

\bibitem[Stacy et al.(2010)]{sta10}
	Stacy, A., Greif, T.~H., \& Bromm, V.\ 2010, \mnras, 403, 45 

\bibitem[Stacy et al.(2012)]{sta12}
	Stacy, A., Greif, T.~H., \& Bromm, V.\ 2012, \mnras, 422, 290 
	
\bibitem[Stone, Mihalas and Norman (1992)]{sto92}
	Stone, J.~M., Mihalas, D., \& Norman, M.~L.\ 1992, \apjs, 80, 819

\bibitem[Tan \& McKee(2004)]{tan04}
	Tan, J.~C., \& McKee, C.~F.\ 2004, \apj, 603, 383 

\bibitem[Tanaka \& Nakamoto(2011)]{tan11}
	Tanaka, K.~E.~I., \& Nakamoto, T.\ 2011, \apjl, 739, L50 

\bibitem[Thompson(2013)]{tho13}
	Thompson, T.~A.\ 2013, \mnras, 431, 63 

\bibitem[Yoshida et al.(2006)]{yos06}
	Yoshida, N., Omukai, K., Hernquist, L., \& Abel, T.\ 2006, \apj, 652, 6

\end{thebibliography}
\end{document}